\documentclass[%
 reprint,
 amsmath,amssymb,
 aps,
]{revtex4-2}

\usepackage{graphicx}
\usepackage{dcolumn}
\usepackage{bm}
\usepackage{hyperref}
\hypersetup{colorlinks, linkcolor={blue}, citecolor={blue}, urlcolor={blue}}

\begin{document}
\title{Longitudinal Spin Hall Magnetoresistance from Spin Fluctuations}

\author{Ping Tang}
\email{tang.ping.a2@tohoku.ac.jp}
\affiliation{WPI-AIMR, Tohoku
University, 2-1-1 Katahira, Sendai 980-8577, Japan}
\date{\today}

\begin{abstract}
Spin Hall magnetoresistance (SMR), the variation in resistance in a heavy metal (HM) with the magnetization orientation of an adjacent ferromagnet (FM), has been extensively studied as a powerful tool for probing surface magnetic moments in a variety of magnetic materials. However, the conventional SMR theory assumes rigid magnetization of a fixed magnitude, an assumption that breaks down close to the FM's Curie temperature \(T_c\), where the magnetic susceptibility diverges. Here, we report an unconventional SMR effect arising from the magnetic-field modulation of spin fluctuations in the FM, while its magnetization remaining collinear to the spin Hall accumulation in the HM. In contrast to the conventional SMR, which scales with the magnetization and vanishes near $T_{c}$, such ``longitudinal" SMR (LSMR), though suppressed at low temperatures, becomes critically enhanced at \(T_c\), reaching a magnitude comparable to conventional SMR amplitudes. Our findings suggest a promising method for electrically detecting enhanced spin fluctuations in magnetic systems.
\end{abstract}

\maketitle
\emph{Introduction.---}Magnetoresistance is a cornerstone of spintronics research, with crucial applications in sensors, data storage, and memory devices \cite{RevModPhys.76.323}, as exemplified by the celebrated giant magnetoresistance \cite{PhysRevLett.61.2472,PhysRevB.39.4828} and tunnel magnetoresistance \cite{julliere1975tunneling,miyazaki1995giant,PhysRevLett.74.3273,yuasa2004giant,parkin2004giant}. Among various MR effects, the spin Hall magnetoresistance (SMR) originally refers to the resistance variation in a heavy metal (HM) with the magnetization direction of an adjacent ferro(ferri)magnet (FM), resulting from the combined action of the (inverse) spin Hall effect and spin transfer torque \cite{PhysRevLett.110.206601,PhysRevB.87.144411,PhysRevB.87.174417,PhysRevB.87.224401,PhysRevB.87.184421,PhysRevB.89.220404,isasa2014spin,meyer2014temperature,cho2015large,PhysRevLett.116.097201,chen2016theory,gomez2020strong}. To date, the SMR effect has been extensively studied as a simple yet powerful tool for detecting the surface magnetic moments of diverse magnetic materials \cite{isasa2016spin,PhysRevB.94.134418,PhysRevApplied.6.034007,lebrun2019anisotropies,feringa2022observation,PhysRevB.106.104426,PhysRevB.108.064434}, including collinear \cite{ ji2017spin,lin2017electrical,hoogeboom2017negative,wang2017antiferromagnetic,PhysRevLett.118.147202,luan2018interfacial,fischer2018spin,ji2018negative,PhysRevB.99.060405,fischer2020large,ross2020structural,PhysRevB.107.054426,sando2024strain} and non-collinear \cite{oda2019magnetoresistance,Mn3Sn} antiferromagnets, non-collinear (canted) ferrimagnets \cite{PhysRevB.94.094401,dong2017spin}, multiferroics \cite{PhysRevB.92.224410,PhysRevMaterials.8.L071401}, and even paramagnets \cite{PhysRevB.98.134402,schlitz2018evolution,lammel2019spin,oyanagi2021paramagnetic,eswara2023spin}. 

The conventional SMR theory assumes the magnetic moments in a magnetic layer to be classical vectors with fixed magnitudes that are unresponsive to an applied magnetic field \cite{PhysRevB.87.144411,chen2016theory,wang2017antiferromagnetic,fischer2018spin}. While this assumption holds at temperatures far below the Curie temperature ($T_{c}$) of the magnetic layer, it becomes questionable near $T_{c}$ at which spin fluctuations are significantly enhanced. SMR experiments on Pt$|$Y\(_3\)Fe\(_5\)O\(_{12}\) (YIG) bilayers at high temperatures have shown that the SMR amplitude scales with net magnetization and decreases rapidly as the temperature approaches the \(T_c\) of YIG \cite{uchida2015spin,wang2015spin}. Recently, a microscopic SMR theory has been developed to account for the effect of spin fluctuations (well below \( T_c \)), suggesting the degradation of the SMR due to magnon excitations \cite{zhang2019theory,PhysRevB.100.180401,kato2020microscopic}. 

\begin{figure}
    \centering
    \includegraphics[width=8.6 cm]{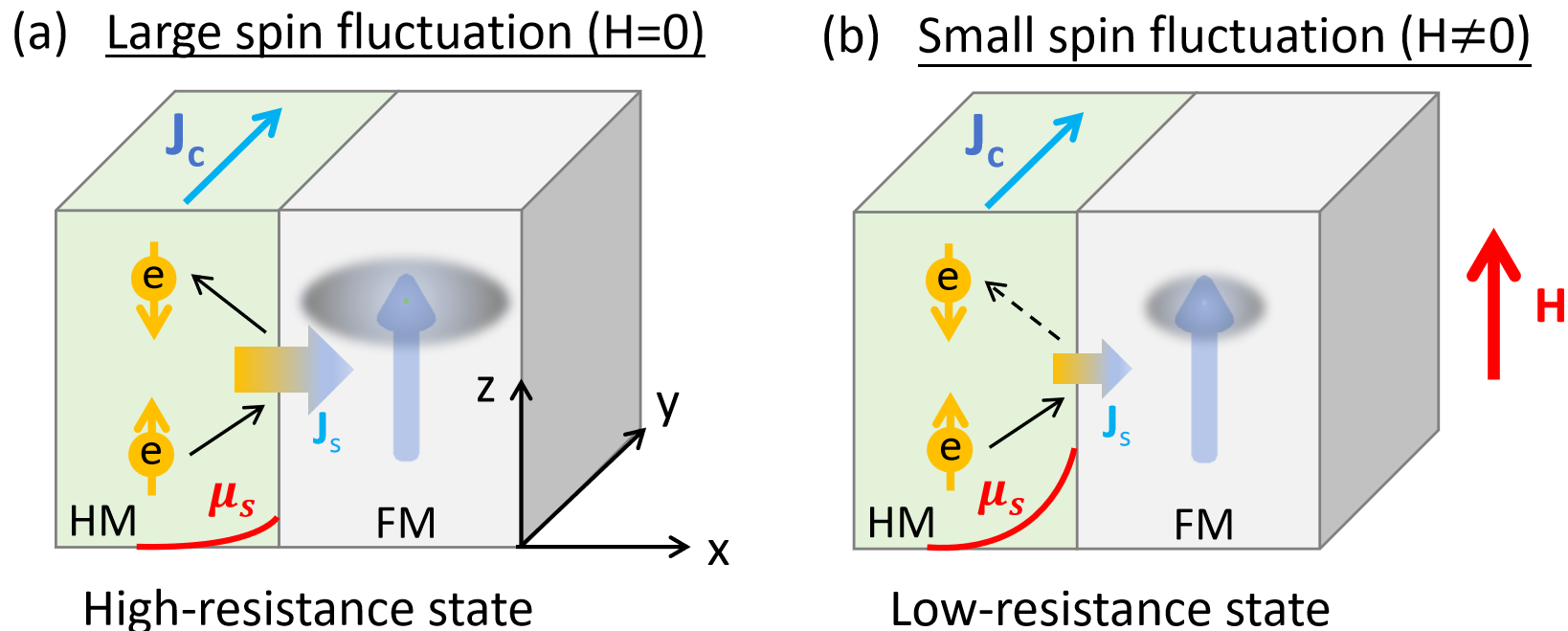}
    \caption{Illustration of the proposed LSMR mechanism in HM$|$FM bilayers with an applied electric current ($J_{c}$) along the $y$ direction. (a) Strong spin (or magnetization) fluctuations in the FM near $T_{c}$ result in high interface spin conductance and reduced electron spin accumulation ($\mu_{s}$) near the interface, which corresponds to a high-resistance state in the HM. (b) An applied magnetic field along the equilibrium magnetization direction of the FM significantly suppresses spin fluctuations and the interface spin conductance, thereby leading to a low-resistance state in the HM. The shadowed cloud around the FM magnetization represents spin fluctuations.}
    \label{Fig-fluctuation}
\end{figure}

In this Letter, we report a previously unrecognized SMR effect in an HM$\vert$FM bilayer, where an applied magnetic field (and thus the magnetization direction) is fixed to align with the spin Hall accumulation in the HM. Unlike the conventional SMR, which arises from changes in magnetization direction under an applied rotating magnetic field, such ``longitudinal" SMR (LSMR) is caused by the magnetic-field modulation of spin fluctuations in the FM layer that alters interface spin conductance, as illustrated in Fig.~\ref{Fig-fluctuation}. We show that, although suppressed at low temperatures, the LSMR ratio, defined as the relative change in the HM resistivity under an applied field [see Eq.~(\ref{LSMR})], is critically enhanced at the \(T_c\) of the FM, in stark contrast to the temperature dependence of the conventional SMR \cite{uchida2015spin,wang2015spin}. Specifically, we calculate an LSMR ratio of $\lesssim10^{-4}$ for an applied moderate magnetic field in a La$_2$NiMnO$_6$$|$Pt bilayer near the $T_{c}$ of La$_2$NiMnO$_6$, an insulating FM with $T_{c}\approx 270\,$K \cite{FMs,PhysRevLett.113.266602}, comparable to the conventional SMR ratio in typical YIG$|$Pt bilayers at room temperature \cite{PhysRevLett.110.206601,PhysRevB.87.174417,PhysRevB.87.224401,PhysRevB.87.184421}. 

\emph{Model.---}We illustrate the LSMR mechanism by focusing on an HM$|$FM bilayer, where the spin Hall accumulation in the HM is polarized collinearly with the FM magnetization or an applied magnetic field (along the $z$ axis) 
 [Fig.~\ref{Fig-fluctuation}]. The interfacial s-d exchange interaction between conduction electrons in the HM and local spins in the FM reads
\begin{align}
\hat{\mathcal{H}}_{sd}=-\sum_{i\in\text{int}}J_{sd}(\mathbf{r}_{i})\hat{\mathbf{S}}_{i}\cdot\hat{\mathbf{s}}_{i}
\label{sdinter}
\end{align}
where $J_{sd}(\mathbf{r}_{i})$ is the sd exchange integral, $\hat{\mathbf{S}}_{i}$ the local spin operator at site $i$ in the FM, and $\hat{\mathbf{s}}_i=(1/2)\sum_{\alpha\beta}c_{i\alpha}^\dag{\boldsymbol\sigma}_{\alpha\beta}c_{i\alpha}$ the spin operator of conduction electrons, with $c_{i\alpha}^{(\dagger)}$ the annihilation (creation) operator of electrons of spin index $\alpha$ and $\sigma_{\alpha\beta}$ the elements of Pauli matrix. The summation is restricted to sites at the HM$|$FM interface. The spin current density flowing into the FM layer is
\begin{align}\label{Is}
J_s= \frac{i}{A_{I}}\sum_{i}\left\langle\left[\hat{s}_i^z,\hat{\mathcal{H}}_{sd}\right]\right\rangle=\frac{1}{A_{I}}\text{Re} \sum_{i\in \text{int}}J_{sd}(\mathbf{r}_{i})\mathcal{G}_{ii}^{<}(t,t)
\end{align}
where $A_{I}$ is the interface area, $\langle \cdots \rangle$ denotes the ensemble average, and $\mathcal{G}_{ii}^{<}(t,t^{\prime})=-i\langle \hat{S}_{i}^-(t^{\prime})\hat{s}_i^+(t)\rangle$ is the lesser Green function that describes correlations of transverse spin fluctuations between the FM and HM layers. Eq.~(\ref{Is}) can be computed using many-body techniques with the sd exchange interaction treated as a perturbation. Here we retain terms in Eq.~(\ref{Is}) up to the second order in $J_{sd}$, which corresponds to
\begin{align}
\mathcal{G}_{ii}^{<}(t,t)=&-\frac{1}{2}\frac{J_{sd}}{\hbar}\sum_{i^{\prime}\in\text{int}}\int dt^{\prime}\left[\chi_{ii^{\prime}}^R(t,t^{\prime})\mathcal{D}_{i^{\prime}i}^{<}(t^{\prime},t)\right.\nonumber\\
&\left.+\chi_{ii^{\prime}}^<(t,t^{\prime})\mathcal{D}_{i^{\prime}i}^{A}(t^{\prime},t)\right]
\end{align}
where $\chi_{ii^{\prime}}^R(t,t^{\prime})=-i\Theta(t-t^{\prime})\langle [\hat{s}_{i}^{+}(t),\hat{s}_{i^{\prime}}^{-}(t^{\prime})]\rangle_{0}$ and $\chi_{ii^{\prime}}^<(t,t^{\prime})=-i\langle \hat{s}_{i^{\prime}}^{-}(t^{\prime})\hat{s}_{i}^{+}(t)\rangle_{0}$ are the transverse retarded and lesser spin susceptibilities of conduction electrons, respectively, while $\mathcal{D}_{ii^{\prime}}^A(t,t^{\prime})=i\Theta(t^{\prime}-t)\langle[\hat{S}_{i}^{+}(t),\hat{S}_{i^{\prime}}^{-}(t^{\prime})]\rangle_0$ and $\mathcal{D}_{ii^{\prime}}^<(t,t^{\prime})=-i\langle\hat{S}_{i^{\prime}}^{-}(t^{\prime})\hat{S}_{i}^+(t)\rangle_0$ the transverse advanced and lesser susceptibilities of localized spins. The notation $\langle\cdots \rangle_{0}$ indicates the average over unperturbed states. Assuming a rough interface with $\overline{J_{sd}(\mathbf{r}_{i})}=0$ and $\overline{J_{sd}(\mathbf{r}_{i})J_{sd}(\mathbf{r}_{j})}=J_{0}^{2}\delta_{ij}$, the spin current density averaged over interface disorder configurations (denoted by the overline) is given by
\begin{align}\label{Is2}
J_s=& -\frac{N_{I}J_{0}^{2}  }{2\hbar A_{I}N_{e}N_{m}}\int\frac{d\omega}{2\pi}\sum_{\mathbf{k}\mathbf{q}}\text{Re}\left[\chi_{\mathbf{q}\omega}^R\mathcal{D}_{\mathbf{k}\omega}^{<}+\chi_{\mathbf{q}\omega}^<\mathcal{D}_{\mathbf{q}\omega}^{A}\right]
\end{align}
in terms of the Fourier components of various spin susceptibilities. Here $N_{e}$ ($N_{m}$) and $N_{I}$ are the number of sites in the HM (FM) and at the interface, respectively. While the generalization to the case of interacting electrons is straightforward, we consider a free (noninteracting)-electron model for the HM with 
\begin{align}
\chi^{R}_{\mathbf{q}\omega}=&\frac{\hbar}{N_{e}}\sum_{\bf k}\frac{f_{{\bf k}\uparrow}-f_{{\bf k}+{\bf q}\downarrow}}{\hbar\omega-\xi_{{\bf k}+{\bf q}}+\xi_{{\bf k}}+i0^+}\label{chi1}\\
\chi^{<}_{\mathbf{q}\omega}=&-\frac{2\pi i\hbar}{N_{e}}\sum_{\bf k}(1-f_{{\bf k}\uparrow})f_{{\bf k}+{\bf q}\downarrow}\delta(\hbar\omega+\xi_{\mathbf{k}}-\xi_{\mathbf{k}+\mathbf{q}}) \label{chi2}
\end{align}
where $f_{{\bf k}\sigma}=\{\exp[{(\varepsilon_{\mathbf{k}}-\mu_{\sigma})/k_{B}T}]+1\}^{-1}$ is the distribution function of spin-$\sigma$ electrons with energy $\varepsilon_{\mathbf{k}}$ and chemical potential $\mu_\sigma$. A non-equilibrium spin accumulation in the HM corresponds to a nonzero spin chemical potential $\mu_{s}=\mu_{\uparrow}-\mu_{\downarrow}$. In the linear response regime, the susceptibilities of local spins in Eq.~(\ref{Is2}) can be replaced with their equilibrium ones that obey the fluctuation-dissipation theorem, $\mathcal{D}_{\mathbf{q}\omega}^{<}=2 i\text{Im} \mathcal{D}_{\mathbf{q}\omega}^{R} n_{\omega}$, where $n_{\omega}=[\exp{(\hbar\omega/k_{B}T)}-1]^{-1}$. Note, however, that a similar fluctuation-dissipation relation does not hold for the non-equilibrium electrons with $\mu_{s}\neq0$. Substituting Eq.~(\ref{chi1}) and Eq.~(\ref{chi2}) into Eq.~(\ref{Is2}) yields the linear-response interface spin conductance:
\begin{align}
G_{\text{int}}\equiv&\frac{J_{s}}{\mu_{s}}=\frac{\pi N_{I}J_{0}^{2}}{2A_{I}N_{e}^{2}N_{m}}\int\frac{d\omega}{2\pi}\sum_{{\bf k}\mathbf{k}^{\prime}{\bf q}}\left\{(n_\omega+f_{{\bf k}^{\prime}}^{0})\frac{\partial f_{\bf k}^{0}}{\partial\xi_{\bf k}}\right.\nonumber\\
&\left.+(n_\omega+1-f_{\bf k}^{0})\frac{\partial f_{\mathbf{k}^{\prime}}^{0}}{\partial\xi_{\mathbf{k}^{\prime}}}\right\}\text{Im} \mathcal{D}_{{\bf q}\omega}^{R}\nonumber\\
&\times\delta(\hbar\omega+\xi_{\bf k}-\xi_{{\bf k}^{\prime}}) \label{Gint}
\end{align}
where $f_{\mathbf{k}}^{0}$ is the equilibrium Fermi-Dirac distribution, and we have used $D_{\mathbf{q}\omega}^{A}=[D_{\mathbf{q}\omega}^{R}]^{\ast}$. Eq.~(\ref{Gint}) closely resembles the low-temperature interface spin conductance derived from sd magnon-electron scattering \cite{takahashi2010spin,PhysRevLett.108.246601,zhang2012}, but it holds for the FM at any temperature, where its spin fluctuations captured by the imaginary part of $\mathcal{D}_{\mathbf{q}\omega}^{R}$ \cite{noteD}. 

\begin{figure}
    \centering
    \includegraphics[width=6cm]{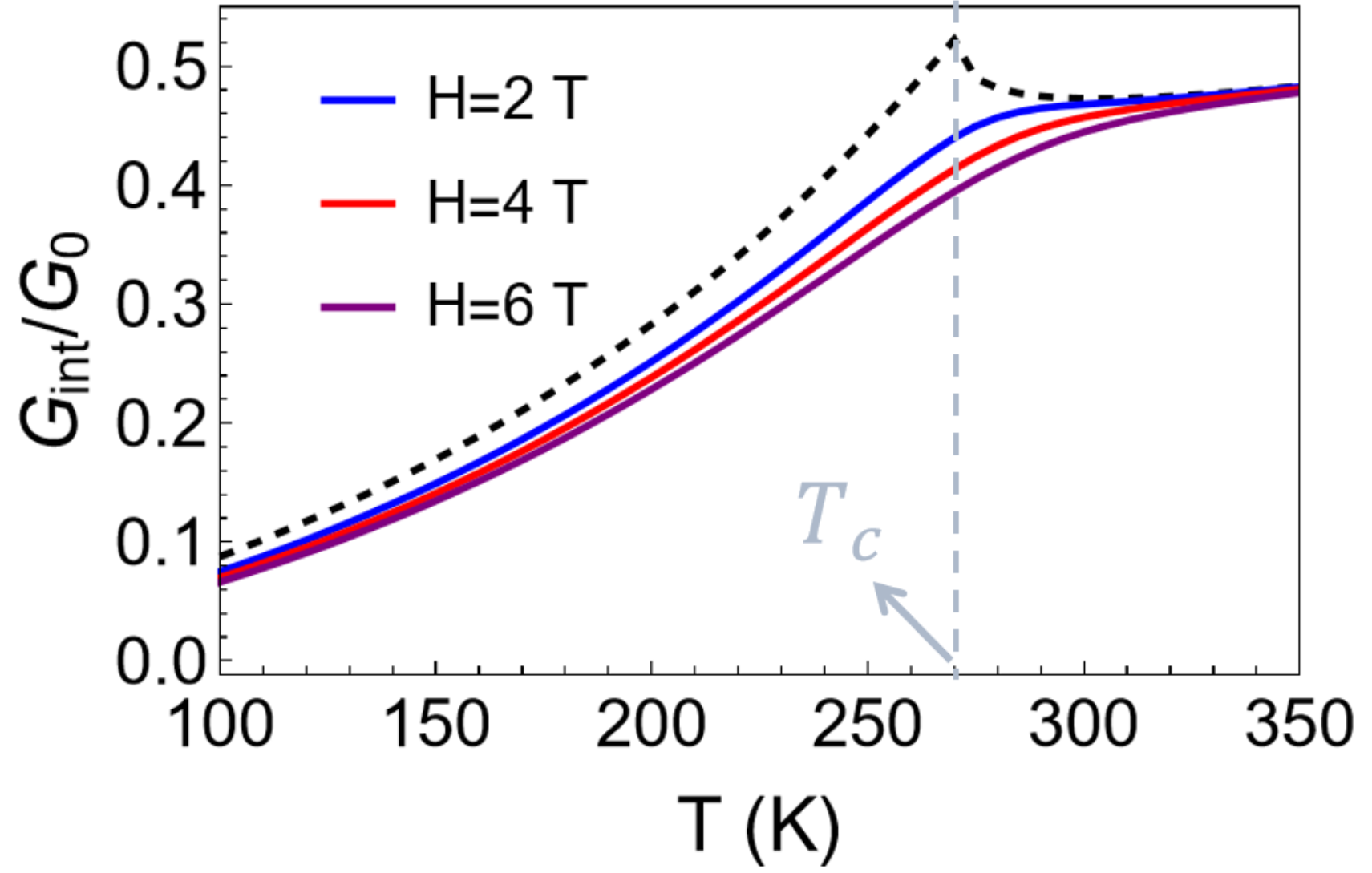}
    \caption{The temperature dependence of the normalized interface spin conductance Eq.~(\ref{Gint2}) at different fields for La$_2$NiMnO$_6$$|$Pt bilayers. The dashed curve corresponds to the case without an applied magnetic field, while the dashed vertical line marks the Curie temperature ($T_{c}\approx 270\,$K) of La$_2$NiMnO$_6$.}
    \label{Fig-Gint}
\end{figure}

Next, we calculate $\mathcal{D}_{\mathbf{q}\omega}^{R}$ across a wide temperature range for a Heisenberg FM in the presence of an applied magnetic field. The spin Hamiltonian of the FM reads
\begin{align}\label{d}
\hat{\mathcal{H}}_{\text{FM}}=&-J\sum_{\langle i,j\rangle}\hat{\bf S}_{i}\cdot\hat{\bf S}_{j}+g\mu_{B}H\sum_{i}\hat{S}_{i}^{z}
\end{align}
where $J$ describes the nearest-neighbor exchange interaction, $g$ the Land\'e factor, and $\mu_{B}$ the Bohr magneton. Since we address the FM at temperatures spanning the $T_{c}$, the standard Holstein-Primakoff expansion of local spin operators into magnon operators fails. Instead, we employ the well-established double-time temperature-dependent Green function approach, which remains valid across the entire temperature range and yields \cite{10.1143/PTP.25.1043,PhysRev.127.88,PhysRev.130.890} 
\begin{equation}
\mathcal{D}_{\mathbf{q}\omega}^{R}=\frac{2\langle \hat{S}_{i}^{z}\rangle}{\omega-\omega_{\mathbf{q}}+i0^{+}}. \label{D}
\end{equation}
Here, $\omega_{\mathbf{q}}(T,H)$ represents the temperature-dependent frequency spectrum of quasi-boson excitations. Within the Tyablikov approximation \cite{10.1143/PTP.25.1043,PhysRev.127.88}, which neglects correlations between longitudinal and transverse fluctuations of local spins, $\omega_{\mathbf{q}}(T,H)=g\mu_{B} H+z_{0}JS m (1-\gamma_{\mathbf{q}})$, where $S$ is the spin quantum number, $m(T,H)=\langle \hat{S}_{i}^{z}\rangle/S$ the reduced magnetization, and $\gamma_{\mathbf{q}}=(1/z_{0})\sum_{\boldsymbol{\delta}_{ij}}\exp(i\mathbf{q}\cdot \boldsymbol{\delta}_{ij})$ the form factor depending on the crystal structure of the FM, with $z_{0}$ the coordination number and $\boldsymbol{\delta}$ the displacement vector connecting nearest-neighbor sites. In the low-temperature limit, $m\rightarrow{1}$ and $\omega_{\mathbf{q}}(T,H)$ reduces to spin wave dispersions, while $\omega_{\mathbf{q}}\rightarrow{0}$ becomes soft as $m\rightarrow{0}$ at $T_{c}$, signaling the magnetic phase transition, analogous to the bosonic excitations of ``ferrons" in ferroelectrics \cite{PhysRevB.106.L081105,PhysRevB.109.L060301,PhysRevApplied.20.050501}. With a self-consistently determined magnetization, the quasi-boson spectrum has been shown to capture the essential spin dynamic properties of a Heisenberg FM over a whole temperature range, despite some quantitative inaccuracies in describing the softening of spin-wave dispersion at low temperatures \cite{PhysRev.130.890}. Substituting Eq.~(\ref{D}) to Eq.~(\ref{Gint}) leads to
\begin{align}
G_{\text{int}}(T,H)
=G_{0}\mathcal{V}\int \frac{d\mathbf{q}}{(2\pi)^{3}}\frac{\hbar\omega_{\mathbf{q}}}{k_{B}T} \frac{me^{\hbar\omega_{\mathbf{q}}/k_{B}T}}{[e^{\hbar\omega_{\mathbf{q}}/k_{B}T}-1]^{2}} \label{Gint2}
\end{align}
where $G_{0}=\pi S N_{I}J_{0}^{2}\mathcal{N}_{F}^{2}/(4A_{I})$ is a temperature-independent interface spin conductance (in units of m$^{-2}$), associated with the zero-temperature spin mixing conductance for spin pumping \cite{PhysRevLett.133.036701}, with $\mathcal{N}_{F}$ the density of state per site in the HM at the Fermi level and $\mathcal{V}$ the volume of magnetic unit cells. The temperature dependence of $G_{\text{int}}$ at different applied fields is shown in Fig.~\ref{Fig-Gint} for the Pt$|$La$_2$NiMnO$_6$ bilayer (studied below). While the magnetic field dependence of the interface spin conductance is negligible at low temperatures, it is significantly enhanced near $T_{c}$, where the magnetic susceptibility $\partial m/\partial H\rightarrow{\infty}$ diverges, implying the relevance of our proposed LSMR mechanism. We note that a similar enhancement of spin pumping into a fluctuating magnetic material near $T_{c}$ has been reported \cite{PhysRevB.89.174417,PhysRevLett.116.186601,PhysRevLett.116.186601,PhysRevB.96.054436}. 

\emph{Results.---}The change in the interface spin conductance alters the HM resistance by modifying the boundary condition for the spin Hall current at the HM$\vert$FM interface. For an applied electric current $J_{c}$ along the $y$-direction, as depicted in Fig.~\ref{Fig-fluctuation}, the electric resistivity in the HM is calculated using the spin drift-diffusion theory as
\begin{equation}
\rho(T, H)=\rho_0\left[1-\theta_{s}^2\left(\frac{\lambda}{d}\right)\frac{4G_N\sinh^2\frac{d}{2\lambda}+G_{\text{int}}\sinh\frac{d}{\lambda}}{G_N\sinh \frac{d}{\lambda}+G_{\text{int}}\cosh \frac{d}{\lambda}}\right] \label{rho}
\end{equation}
where $\rho_{0}$, $\theta_{s}$, $d$, $\lambda$, and $G_N\equiv\hbar/(4e^{2}) (\lambda\rho_{0})^{-1}$ are the bulk electric resistivity, spin Hall angle, thickness, spin diffusion length, and bulk spin conductance of the HM layer, respectively. In contrast to the conventional SMR, which depends on the magnetic field direction, we define the LSMR ratio
\begin{equation}
\text{LSMR}\equiv\frac{\rho(T,0)-\rho(T,H)}{\rho_{0}} \label{LSMR}
\end{equation}
which varies with the field \emph{magnitude} through $G_{\text{int}}(T,H)$. Here \text{LSMR} is positive, as the applied field reduces $G_{\text{int}}$ by suppressing spin fluctuations in the FM. 
\begin{figure}
    \centering
    \includegraphics[width=6cm]{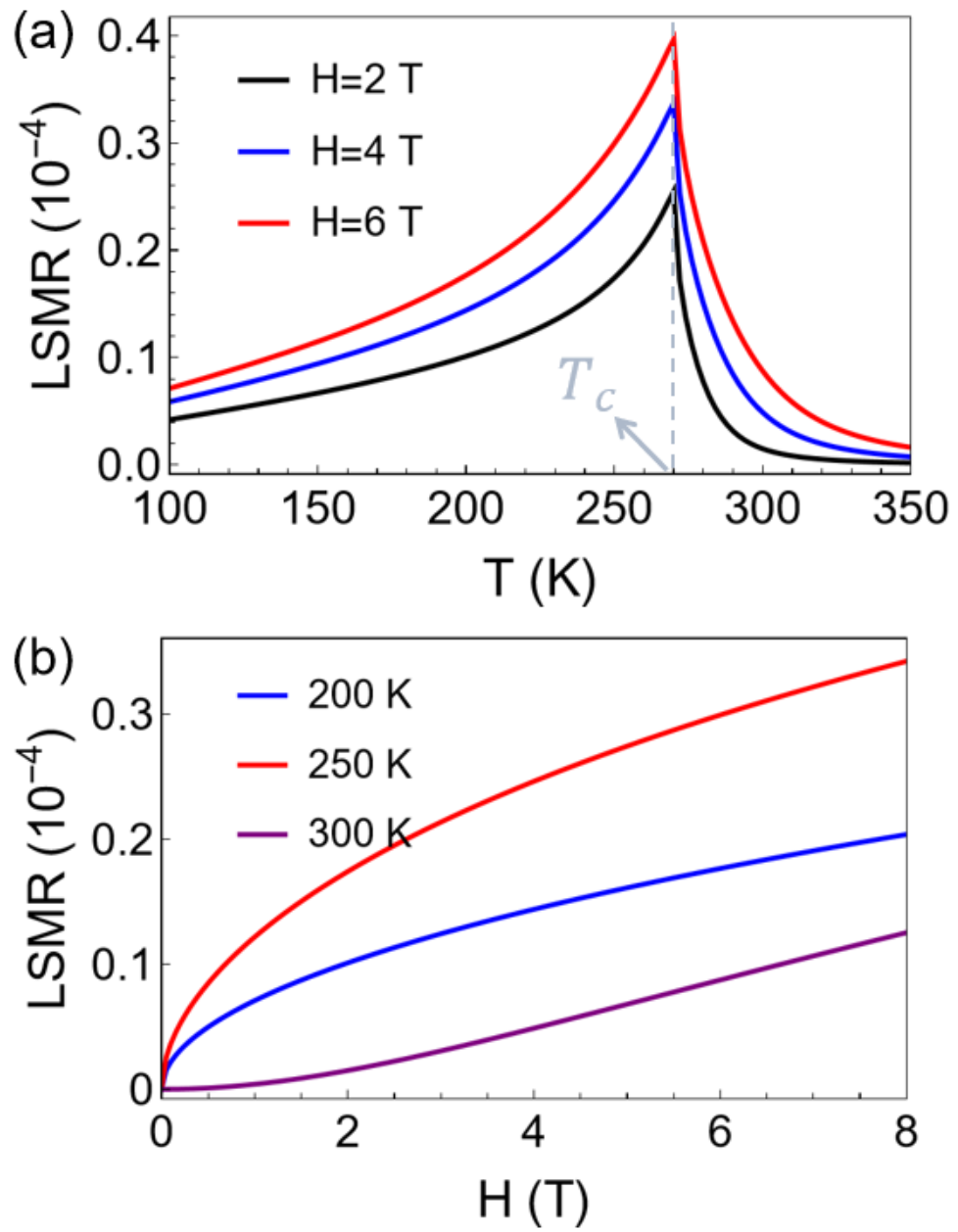}
    \caption{The LSMR in La\(_2\)NiMnO\(_6\)\(|\)Pt bilayers as a function of (a) temperature and (b) the applied magnetic-field magnitude.}
    \label{fig-LSMR}
\end{figure}

We present numerical calculations based on Eq.~(\ref{Gint2}) and Eq.~(\ref{rho}) for the La$_2$NiMnO$_6$$|$Pt bilayer, in which paramagnetic spin pumping was previously observed \cite{PhysRevLett.113.266602}. La$_2$NiMnO$_6$ is a double-perovskite insulating FM with a $T_{c}$ close to room temperature; each formula cell contains one Ni$^{2+}$ ($S=1$) spin and one Mn$^{4+}$ ($S=3/2$) spin, which are ferromagnetically coupled by the superexchange interaction. To avoid inessential complications, we model $m(T)$ for La$_2$NiMnO$_6$ using the molecular-field theory of a single-sublattice FM with an effective spin number $\bar{S}=5/2$, i.e., $m=B_{\bar{S}}[g\mu_{B} \bar{S}(H+mH_{M})/(k_{B}T)]$, where $B_{\bar{S}}(x)$ is the Brillouin function and $H_{M}$ a temperature-independent molecular field. We set $g\approx 2$ and $H_{M}=172\,$T to correctly reproduces $T_{c}=270\,$K of La$_2$NiMnO$_6$. In Eq.~(\ref{Gint2}), the quasi-boson dispersion is approximated by $\omega_{\mathbf{q}}(T,H)=g\mu_{B} H+ m D q^2$, with $D$ the zero-temperature exchange stiffness, and the integral over $\mathbf{q}$ is cut off at $q_{c}=(6\pi^{2}/\mathcal{V})^{1/3}$, chosen so that $\mathcal{V}/(2\pi)^{3}\int d\mathbf{q}=1$. Fig.~\ref{fig-LSMR} presents the temperature and magnetic field dependence of the LSMR ratio, where model parameters are taken as $D =2\times 10^{-6}\,$m$^2$/s, $G_{0}=10^{17}\,$m$^{-2}$, $\mathcal{V}=58\,$\AA$^{3}$ \cite{NoteV}, $\rho_{0}^{-1}=2\times10^{6}\,$S/m, $d=5\,$nm, $\lambda =2\,$nm, and $\theta_{s}=0.11$. Fig.~\ref{fig-LSMR}(a) shows that below $T_{c}$, the LSMR ratio increases rapidly with temperature but then drops sharply above $T_{c}$, peaking at a magnitude of $\lesssim 10^{-4}$ at $T_{c}$, which is comparable to the conventional SMR amplitudes in YIG$|$Pt bilayers \cite{PhysRevLett.110.206601,PhysRevB.87.174417,PhysRevB.87.224401,PhysRevB.87.184421}. Remarkably, the LSMR does not vanish abruptly above $T_{c}$, since the field-driven modification of paramagnetic spin fluctuations remains pronounced near $T_c$ because of the large paramagnetic susceptibility, whereas the conventional paramagnetic SMR arises from the field-induced magnetization \cite{PhysRevB.98.134402,schlitz2018evolution,lammel2019spin,oyanagi2021paramagnetic,eswara2023spin}. Fig.~\ref{fig-LSMR}(b) plots the LSMR as a function of the magnetic field magnitude at several temperatures. In contrast to the conventional SMR, which relies only on the magnetization (field) direction, the LSMR scales with the field amplitude. These distinct temperature- and field-dependent behaviors of the LSMR and conventional SMR suggest a sign change in the \emph{measured} SMR amplitude at a certain temperature, above which the LSMR exceeds the conventional SMR, i.e., the resistance in the HM$|$FM is high (low) when the spin Hall accumulation is parallel (perpendicular) to the FM magnetization. Indeed, a sign change of the SMR with temperature has been reported, seemingly indicating the LSMR mechanism \cite{PhysRevB.94.094401,dong2017spin,PhysRevLett.118.147202,kato2020microscopic}. However, a comprehensive comparison of the LSMR and conventional SMR across the full temperature range and varying magnetic-field magnitudes is beyond the scope of the present work.

\emph{Conclusions.---}In conclusion, we have proposed an unconventional LSMR in HM$|$FM bilayers driven by the magnetic field modulation of spin fluctuations in the FM. Unlike the conventional SMR, which relies on magnetic ordering and diminishes with increasing temperature, the LSMR is critically enhanced near the FM's Curie temperature, both above and below, and can surpass the conventional SMR. While the conventional SMR is widely used to reveal static magnetic structures in spin-ordered systems, our proposed LSMR offers a promising electrical means to probe enhanced spin fluctuations in a broad range of materials. These are not limited to conventional magnets near the phase transition but also include other (para)magnetic materials, such as spin liquids \cite{savary2016quantum,zhou2017quantum}, spin-frustrated systems \cite{diep2013frustrated}, and low-dimensional magnets \cite{PhysRevLett.17.1133}, opening new avenues for exploring spin fluctuation-driven phenomena in exotic magnetic states. Our work might also shed light on the ``nonlocal" magnetoresistance recently observed in YIG$|$NiO$|$YIG$|$Pt heterostructures \cite{guo2020nonlocal}, in which the parallel (antiparallel) magnetization configuration of two YIG layers corresponds to a low (high)-resistance state in Pt, in terms of enhanced spin fluctuations in the antiferromagnetic configuration. 

\textit{Acknowledges.---}The author thanks Gerrit E. W. Bauer for valuable suggestions. The author also acknowledges financial support from JSPS KAKENHI Grant-in-Aid for Scientific Research (S) (No. 22H04965) and Grant-in-Aid for Early-Career Scientists (No. 23K13050).

\bibliography{reference}

\begin{thebibliography}{78}
\bibitem [{not()}]{noteD}%
  \BibitemOpen
  \href@noop {} {}\bibinfo {note} {{At low temperatures $D_{\mathbf{q}\omega}^{R}$ reduces to the magnon retarded Green function and its imaginary part gives the ferromagnetic resonance absorption}}\BibitemShut {NoStop}%
\bibitem [{Not()}]{NoteV}%
  \BibitemOpen
  \href@noop {} {}\bibinfo {note} {{ Here the volume of a formula cell in \text{La$_2$NiMnO$_6$} is adopted, which contains one Ni$^{2+}$ and one Mn$^{4+}$}}\BibitemShut {NoStop}%
\end{thebibliography}

\end{document}